\documentclass[prb,floatfix,superscriptaddress,showpacs]{revtex4}
\usepackage{epsfig}

\newcommand{\be}{\begin{equation}}
\newcommand{\ee}{\end{equation}}

\newcounter{appendix}
\renewcommand{\theappendix}{\Alph{appendix}.}

\begin{document}

\title
{{\em Ab Initio} calculation of band gap renormalization in highly
excited GaAs}
\author{Catalin D. Spataru} 
\affiliation{Department of Physics, University of California at Berkeley,
Berkeley, California 94720}
\affiliation{Materials Sciences Division, Lawrence Berkeley National Laboratory, Berkeley, California 94720}
\author{Lorin X. Benedict}
\affiliation{H Division, Physics and Advanced Technologies Directorate, Lawrence Livermore National Laboratory, University of California, Livermore, CA 94550}
\author{Steven G. Louie}
\affiliation{Department of Physics, University of California at Berkeley,
Berkeley, California 94720}
\affiliation{Materials Sciences Division, Lawrence Berkeley National
Laboratory, Berkeley, California 94720}

\date{\today}

\begin{abstract}
We present {\em ab initio} quasiparticle self-energy calculations in crystalline
GaAs for cases of intense electronic excitation ($\sim 10\%$ of valence
electrons excited into conduction band), relevant for high-intensity
ultra-short pulsed laser experiments. Calculations are performed using
an out-of-equilibrium generalization of the $GW$ approximation based on the
Keldysh Green's function approach. Our results indicate that while the
quasiparticle band
gap is a sensitive function of the amount of excitation, it is not
possible to induce complete band gap closure in this system by purely
electronic means.
\end{abstract}
\maketitle

\section{Introduction}
The ability to optically pump semiconductors with ultra-short pulsed lasers 
has led to the discovery of a host of phenomena over the past few decades 
\cite{Chemla}. One of the most interesting is band gap renormalization 
(BGR), or a change in the quasiparticle gap of a material which occurs when 
electrons are redistributed in energy. The classic scenario is one in which 
electrons are excited from the top of the valence band to the bottom of the 
conduction band (in a direct-gap material such as GaAs), leaving holes 
behind. Though measurements of this effect are often difficult, reductions
in the band gap have been inferred in the $\sim$ 10 meV range for excited 
carrier densities of $\sim$ 10$^{15}$ - 10$^{16}$ $\frac{e^{-}}{cm^{3}}$ 
\cite{Chemla,excuse_BGR}. Since these carrier densities are relatively 
small, the material remains crystalline during the duration of an optical 
pumping experiment. When the carriers have relaxed to the extrema of the 
bands, they are amenable to theoretical treatment 
within the effective mass approximation. The gap reduction itself is 
determined by calculating the quasiparticle self-energy in a 
screened-interaction (e.g. $GW$ approximation \cite{GW1,GW2}) perturbation 
theoretic treatment \cite{excuse_BGR,SR}. 
Extensions of these approaches which utilize {\em ab initio} electronic 
structure techniques have been applied to BGR in conditions of high doping 
levels, in which excess electrons (but no holes) are present \cite{Godby}. 

In the last few years, optical pump-probe experiments have been performed 
which reach carrier densities (electrons + holes) in excess of 
$\sim$ 10$^{21}$ $\frac{e^{-}}{cm^{3}}$ \cite{Mazur1}. In these experiments, 
intense short-pulsed laser irradiation of GaAs rapidly melts the material 
and creates a metallic amorphous state \cite{Allen}. Though the non-thermal 
melting is a primary goal of the studies, significant amounts of optical 
data were recorded prior to the time at which melting is thought to 
occur. For these short pump-probe delay times ($\sim$ few hundred fs), 
deviations of the dielectric function from its values in the unpumped state 
are expected to be mostly electronic in origin (the heavier ions still being 
close to their equilibrium lattice positions). One of us (L.X.B.) has 
studied the changes in optical properties that result from the blocking of 
transitions (Pauli blocking) and the unbinding of exciton states which may 
occur in such conditions \cite{Benedict}. However in that work, BGR was 
neglected. 

Calculations of BGR in GaAs for these conditions of excitation have been 
performed by Kim et al. \cite{Kim} using an empirical pseudopotential technique together 
with a quasistatic effective mass-derived self-energy contribution. Their 
conclusion was that complete band gap closure would occur if $\sim$ 10$\%$ 
of the electrons are excited from the top of the valence band to the bottom 
of the conduction band. In their theory, a significant portion of the band 
gap reduction is due to the change in the screened electron-ion potential 
that results from the excitation of free carriers, the rest being caused by 
the changes in the quasiparticle self-energy. The change in Hartree 
potential (the interaction between electrons at the mean-field level) is 
neglected in their approach. 

In this work, we calculate BGR for laser-excited GaAs in which 
electrons have been excited from the top of the valence band to occupy
the lowest energy states of the bottom of 
the conduction band. Single-particle states are taken from {\em ab initio} 
pseudopotential local density approximation (LDA) band structure calculations, 
and the quasiparticle self-energy is computed in the $GW$ approximation 
(which includes dynamical screening of the electron-electron
interaction) \cite{GW2}. 
The LDA calculations are done for constrained (excited) occupation of the 
electronic states, so the self-consistent contribution of the Hartree
term (owing to the redistribution of charge density) is 
included. We find, contrary to the results of Kim et al. \cite{Kim}, that it 
is impossible to induce complete band gap closure by purely electronic means, 
assuming that the excitation is of the type described above. The reason is 
that the change resulting from the electrostatic (electron-ion + Hartree) 
terms causes the gap to increase as electrons are moved from the valence band 
top (near As) to the conduction band bottom (near Ga). While our 
{\em ab initio} calculation of the self-energy contribution yields similar 
results to those of the more approximate theories \cite{excuse_BGR,SR}, it 
is the electrostatic contribution which dominates the picture at high carrier 
density ($>$ 10$\%$). In what follows, we describe the derivation of 
the self-energy contribution within the $GW$ approximation using the 
Keldysh technique, present the details of our computation of BGR, and discuss 
our results. 

\section{Theory and Computation}
\subsection{Quasiparticle Self-energy Operator in the GW Approximation:
Non-equilibrium Case}

\begin{figure}
\epsfig{file=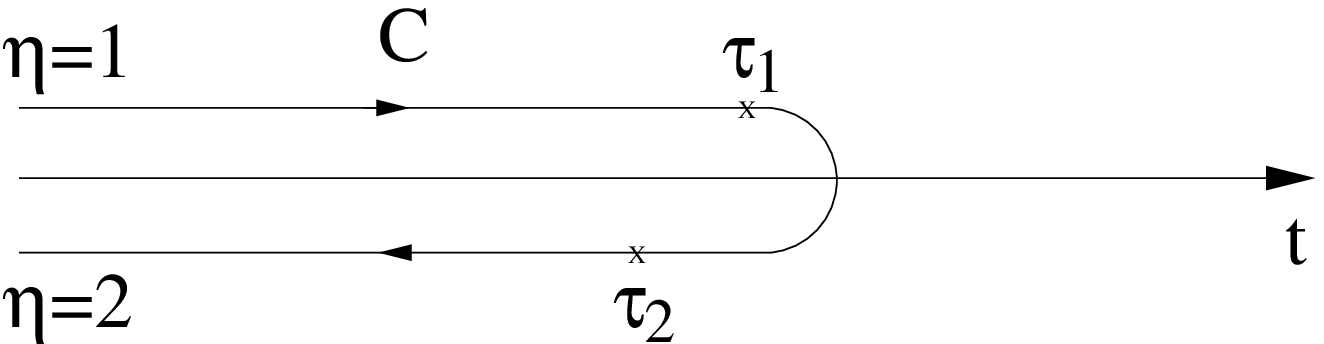, width=10.0cm}
\caption{Closed time path contour C along real axis, showing two
branches, $\eta = 1,2$.}
\label{Fig01}
\end{figure}

We are interested in describing the single-particle excitations of a system 
of interacting electrons out of equilibrium \cite{excuse_single}. 
Thus, we appeal to non-equilibrium quantum many-body theory, in particular, 
the Keldysh method. Using the Keldysh technique \cite{Keldysh}, Green's 
functions, self-energies, etc., for non-equilibrium problems can be constructed 
in a way similar to that for equilibrium problems. The only difference is the 
replacement of time-ordered Green's functions with contour-ordered Green's 
functions \cite{Rammer}:
\begin{equation}
G({\bf r}_1,{\bf r}_2; \tau_1,\tau_2) = -i \langle T_C[\Psi_H({\bf r}_1; \tau_1)
\Psi_H^+({\bf r}_2; \tau_2)] \rangle,
\end{equation}
where the time-labels lie on a contour with two branches: $\tau = t, \eta$ 
($\eta = 1,2$). (See Fig. \ref{Fig01}.) The contour C runs along the real axis, 
starting from -$\infty$
on branch  $\eta=1$, passing through $\tau_1$ and $\tau_2$ once and returning 
to -$\infty$ on branch $\eta=2$. Such a contour allows us to conduct the 
thought-experiment in which the system begins in some excited state of the 
non-interacting system, then the interactions are turned on adiabatically, 
turned slowly off again, and the system is returned to its initial state. This 
intellectual construct is necessary when relating expectation values of 
quantities in an interacting system to the corresponding expectation values 
of the non-interacting system \cite{noneq}. In what follows, we omit the 
spatial variables for simplicity, which are to be used exactly as in
the equilibrium case. We return to the full notation later.

The contour-ordered Green's function obeys the same Dyson equation as
the time-ordered Green's function, if one replaces the real time axis integrals by contour integrals:
\begin{eqnarray}
G(\tau_1,\tau_2) = G_0(\tau_1,\tau_2) + 
\int_{C}d\tau_3 \int_{C}d\tau_4 G_0(\tau_1,\tau_3) \Sigma(\tau_3,\tau_4) G(\tau_4,\tau_2).
\end{eqnarray}

We use the $GW$ approximation for the self-energy:
\begin{eqnarray}
 \Sigma(\tau_1,\tau_2) = i G(\tau_1,\tau_2) W(\tau_1,\tau_2),
\end{eqnarray}
where the screened Coulomb potential $W$ obeys the following
self-consistent equation:
\begin{eqnarray}
W(\tau_1,\tau_2) = V\delta(\tau_1-\tau_2) + V\int_C d\tau_3 P(\tau_1,\tau_3) 
W(\tau_3,\tau_2),
\end{eqnarray}
where $V$ is the Coulomb potential. The polarization function is calculated 
within the random phase approximation (RPA):
\begin{eqnarray}
P(\tau_1,\tau_2) = -i G(\tau_1,\tau_2) G(\tau_2,\tau_1).
\end{eqnarray}

In order to make the calculations tractable, we replace the 
contour integrals by real time integrals. This is achieved
using the Langreth rules for analytic continuation \cite{Langreth},
which we present in the Appendix, after introducing some commonly used notation 
for functions, $A$, defined on the contour (called 'lesser', 'greater', 
'retarded' and 'advanced' respectively): 
\begin{eqnarray}
A^<(t,t') \equiv A(t,\eta=1,t',\eta'=2) 
\nonumber \\
A^>(t,t') \equiv A(t,\eta=2,t',\eta'=1) 
\nonumber \\
A^r(t,t') \equiv \theta(t-t') [A^>(t,t') - A^<(t,t')]
\nonumber \\
A^a(t,t') \equiv \theta(t'-t) [A^<(t,t') - A^>(t,t')].  
\end{eqnarray}
Our goal is now to evaluate the retarded quasiparticle self-energy operator, 
$\Sigma^r$, in the $GW$ approximation. 
From (3), (6), and (A.4) (see Appendix), $\Sigma^r$ becomes:
\begin{eqnarray}
\Sigma^r(t,t')=i[G^<(t,t')W^r(t,t')+G^r(t,t')
W^<(t,t')+G^r(t,t')W^r(t,t')]\nonumber,
\end{eqnarray}
which can be further simplified using (A.1) and (A.2):
\begin{eqnarray}
\Sigma^r(t,t')=i[G^<(t,t')W^r(t,t')+G^r(t,t')W^>(t,t')].
\end{eqnarray}

The Keldysh components of the screened Coulomb interaction are found
using the Langreth rule (A.3):
\begin{eqnarray}
W^r(t,t')= V\delta(t-t') +V \int_{-\infty}^{\infty} dt_1 P^r(t,t_1)W^r(t_1,t') \nonumber \\
W^a(t,t')= V\delta(t-t') +V \int_{-\infty}^{\infty} dt_1 P^a(t,t_1)W^a(t_1,t') \nonumber \\
W^<(t,t')= V \int_{-\infty}^{\infty} dt_1 [P^r(t,t_1)W^<(t'_1,t') +
P^<(t,t_1)W^a(t_1,t')] 
\nonumber \\
W^>(t,t')= V \int_{-\infty}^{\infty} dt_1 [P^r(t,t_1)W^>(t'_1,t') + P^>(t,t_1)W^a(t_1,t')]. 
\end{eqnarray}

It can be shown \cite{Haug,Hartmann} that the
self-consistent equations (8) lead to the following relations for 
the ``greater'' and ``lesser'' components of the screened interaction:
\begin{eqnarray}
W^<(t,t')= \int_{-\infty}^{\infty} dt_1 \int_{-\infty}^{\infty} dt_2 
 W^r(t,t_1)P^<(t_1,t_2)W^a(t_2,t')  
\nonumber \\
W^>(t,t')= \int_{-\infty}^{\infty} dt_1 \int_{-\infty}^{\infty} dt_2 
 W^r(t,t_1)P^>(t_1,t_2)W^a(t_2,t').
\end{eqnarray}

We construct ``non-interacting'' Green's
functions within the quasiparticle approximation from LDA \cite{Sham}
Bloch wavefunctions $|n{\bf k}\rangle$, and eigenvalues $E_{n{\bf
k}}$. Using (1) and (6) we find:

\begin{eqnarray}
\langle n{\bf k}|G^r(\omega)|m{\bf
k}'\rangle=\frac{1}{\omega-E_{n{\bf k}}+i\eta} 
\delta_{nm} \delta_{{\bf k}{\bf k}'}
\nonumber \\
\langle n{\bf k}|G^a(\omega)|m{\bf
k}'\rangle=\frac{1}{\omega-E_{n{\bf k}}-i\eta} \delta_{nm} \delta_{{\bf k}{\bf k}'}
\nonumber \\
\langle n{\bf k}|G^<(\omega)|m{\bf
k}'\rangle=2 \pi i f(E_{n{\bf k}})
\delta(\omega-E_{n{\bf k}})
\delta_{nm} \delta_{{\bf k}{\bf k}'}
\nonumber \\
\langle n{\bf k}|G^>(\omega)|m{\bf 
k}'\rangle=-2 \pi i [1- f(E_{n{\bf k}})]
\delta(\omega-E_{n{\bf k}})
\delta_{nm} \delta_{{\bf k}{\bf k}'},
\end{eqnarray}
where $\eta=0^+$ and we assumed that the ``non-interacting'' system can be
characterized by a  distribution function in energy, $f(E)$.

In frequency space, the expression for the retarded self-energy (7) becomes:
\begin{eqnarray}
\langle n{\bf k}|\Sigma^r(\omega)|n{\bf k}\rangle= i 
\sum_{m,{\bf q},{\bf G}, {\bf G'}}
M_{\bf G}^*(n,m,{\bf k},{\bf -q}) M_{\bf G'}(n,m,{\bf k},{\bf -q}) 
\nonumber \\
\times \int_{-\infty}^{\infty} \frac {d\omega}{2\pi}
[ W^r_{{\bf G}, {\bf G'}}({\bf q},\omega) 
G^<({\bf k}-{\bf q},E_{n{\bf k}}-\omega) + 
W^>_{{\bf G}, {\bf G'}}({\bf q},\omega) 
G^r({\bf k}-{\bf q},E_{n{\bf k}}-\omega)],
\end{eqnarray}
where the matrix elements $M$ are defined by:
$M_{\bf G}(n,m,{\bf k},-{\bf q})=\langle m{\bf k}-{\bf q}|e^
{-i({\bf q}+{\bf G}){\bf r}}|n{\bf k}\rangle$.
Next we need the Fourier components of the screened Coulomb
interaction. Using (8) and (9), the Fourier components of the screened 
Coulomb interaction can be written:
\begin{eqnarray}
W^r_{{\bf G},{\bf G'}}({\bf q},\omega)=[\epsilon^r({\bf 
q},\omega)^{-1}]_{{\bf G},{\bf G'}} V({\bf q}+{\bf G'}) 
\nonumber \\
W^a_{{\bf G},{\bf G'}}({\bf q},\omega)=[\epsilon^a({\bf
q},\omega)^{-1}]_{{\bf G},{\bf G'}} V({\bf q}+{\bf G'})
\nonumber \\
W^<_{{\bf G},{\bf G'}}({\bf q},\omega)= -
[\epsilon^r({\bf q},\omega)^{-1}
\epsilon^<({\bf q},\omega)
\epsilon^a({\bf q},\omega)^{-1}]_{{\bf G},{\bf G'}}
V({\bf q}+{\bf G'})
\nonumber \\
W^>_{{\bf G},{\bf G'}}({\bf q},\omega)= -
[\epsilon^r({\bf q},\omega)^{-1}
\epsilon^>({\bf q},\omega)
\epsilon^a({\bf q},\omega)^{-1}]_{{\bf G},{\bf G'}}
V({\bf q}+{\bf G'}),
\end{eqnarray}
where the dielectric function is defined by:
\begin{eqnarray}
\epsilon^r_{{\bf G},{\bf G'}}({\bf q},\omega)=\delta_{{\bf G},{\bf
G'}} - V({\bf q}+{\bf G})P^r_{{\bf G},{\bf G'}}({\bf q},\omega)
\nonumber \\
\epsilon^a_{{\bf G},{\bf G'}}({\bf q},\omega)=\delta_{{\bf G},{\bf
G'}} - V({\bf q}+{\bf G})P^a_{{\bf G},{\bf G'}}({\bf q},\omega)
\nonumber \\
\epsilon^<_{{\bf G},{\bf G'}}({\bf q},\omega)= - V({\bf q}+{\bf G})
P^<_{{\bf G},{\bf G'}}({\bf q},\omega)
\nonumber \\
\epsilon^>_{{\bf G},{\bf G'}}({\bf q},\omega)= - V({\bf q}+{\bf G})
P^>_{{\bf G},{\bf G'}}({\bf q},\omega).
\end{eqnarray}
The Fourier components of the  polarizability are found via (5), (A.5)
and (10):
\begin{eqnarray}
P^r_{{\bf G},{\bf G'}}({\bf q},\omega)=\sum_{n,m,{\bf p}} 
M_{\bf -G}^*(n,m,{\bf p},{\bf q}) M_{\bf -G'}(n,m,{\bf p},{\bf q})
\frac{f(E_{n{{\bf p}+{\bf q}}})-f(E_{m{\bf
p}})}{E_{n{{\bf p}+{\bf q}}}-\omega-E_{m{\bf p}}-i\eta}
\nonumber \\
P^a_{{\bf G},{\bf G'}}({\bf q},\omega)=\sum_{n,m,{\bf p}} 
M_{\bf -G}^*(n,m,{\bf p},{\bf q}) M_{\bf -G'}(n,m,{\bf p},{\bf q})
\frac{f(E_{n{{\bf p}+{\bf q}}})-f(E_{m{\bf
p}})}{E_{n{{\bf p}+{\bf q}}}-\omega-E_{m{\bf p}}+i\eta} 
\nonumber \\
P^<_{{\bf G},{\bf G'}}({\bf q},\omega)=-2\pi i \sum_{n,m,{\bf p}} 
M_{\bf -G}^*(n,m,{\bf p},{\bf q}) M_{\bf -G'}(n,m,{\bf p},{\bf q})
f(E_{n{{\bf p}+{\bf q}}}) [1-f(E_{m{\bf
p}})] \delta (E_{n{{\bf p}+{\bf q}}}-\omega-E_{m{\bf p}}) 
\nonumber \\
P^>_{{\bf G},{\bf G'}}({\bf q},\omega)=-2\pi i \sum_{n,m,{\bf p}} 
M_{\bf -G}^*(n,m,{\bf p},{\bf q}) M_{\bf -G'}(n,m,{\bf p},{\bf q})
[1-f(E_{n{{\bf p}+{\bf q}}})] f(E_{m{\bf
p}}) \delta (E_{n{{\bf p}+{\bf q}}}-\omega-E_{m{\bf p}}). 
\end{eqnarray}
Decomposing the self-energy into the screened-exchange and
Coulomb-hole parts \cite{GW2}:
\begin{eqnarray}
\langle n{\bf k}|\Sigma^r(E)|n{\bf k}\rangle = 
\langle n{\bf k}|\Sigma^r_{SX}(E)|n{\bf k}\rangle 
+ 
\langle n{\bf k}|\Sigma^r_{CH}(E)|n{\bf k}\rangle, \nonumber
\end{eqnarray}
the final expression for the retarded self-energy is:
\begin{eqnarray}
\langle n{\bf k}|\Sigma^r_{SX}(E)|n{\bf k}\rangle = - \sum_{m,{\bf q},{\bf G}, {\bf G'}}
M_{\bf G}^*(n,m,{\bf k},{\bf -q}) M_{\bf G'}(n,m,{\bf k},{\bf -q})
\nonumber \\ \times
[\epsilon^r({\bf q},E-E_{m{\bf k}-{\bf q}})^{-1}]_
{{\bf G},{\bf G'}} V({\bf q}+{\bf G'}) f(E_{m{\bf k}-{\bf q}}) \\
\langle n{\bf k}|\Sigma^r_{CH}(E)|n{\bf k}\rangle = - \sum_{m,{\bf q},{\bf G}, {\bf G'}}
M_{\bf G}^*(n,m,{\bf k},{\bf -q}) M_{\bf G'}(n,m,{\bf k},{\bf -q})
\nonumber \\ \times
i \int_{-\infty}^{\infty} \frac {d\omega}{2\pi} 
[\epsilon^r({\bf q},\omega)^{-1}
\epsilon^>({\bf q},\omega)
\epsilon^a({\bf q},\omega)^{-1}]_{{\bf G},{\bf G'}}
\frac{1}{E-\omega-E_{m{\bf k}-{\bf q}}+i\eta}
V({\bf q}+{\bf G'}).
\end{eqnarray}

The above result represents a general expression for the self-energy 
(within the $GW$ approximation) of a system characterized by an arbitrary 
distribution function, $f(E)$.
In particular, in the case of a system in equilibrium at some 
temperature, $f(E)$ is just the 
Fermi distribution function, $f_F$, and the following holds via (13) and (14):
\begin{eqnarray}
\epsilon^>_{{\bf G},{\bf G'}}({\bf q},\omega)=
\epsilon^<_{{\bf G},{\bf G'}}({\bf q},\omega)[1+\frac{1}{f_B(\omega)}],
\nonumber
\end{eqnarray}
where $f_B$ denotes the Bose distribution function. Using this result 
together with the Keldysh relation (A.1) for $\epsilon$ , it can be
shown that (16) becomes:
\begin{eqnarray}
\langle n{\bf k}|\Sigma^r_{CH}(E)|n{\bf k}\rangle = - \sum_{m,{\bf q},{\bf G}, {\bf G'}}
M_{\bf G}^*(n,m,{\bf k},{\bf -q}) M_{\bf G'}(n,m,{\bf k},{\bf -q})
\nonumber \\ \times
\frac {1}{\pi}\int_{-\infty}^{\infty} d\omega 
\frac {[\epsilon^r({\bf q},\omega)^{-1}
       -\epsilon^a({\bf q},\omega)^{-1}]_{{\bf G},{\bf G'}}}{2i}
\frac{1+f_B(\omega)}{E-\omega-E_{m{\bf k}-{\bf q}}+i\eta}
V({\bf q}+{\bf G'}).
\end{eqnarray}
For systems with inversion symmetry, we can further replace in (17) 
$\frac {[\epsilon^r({\bf q},\omega)^{-1}-
\epsilon^a({\bf q},\omega)^{-1}]}{2i}$ 
by $Im [\epsilon^r({\bf q},\omega)^{-1}]$
and thus we recover a previous result \cite{Benedict3}, 
obtained by applying the finite
temperature Matsubara Green's function formalism to a system 
with inversion symmetry.

\subsection{Computational details: GaAs}

We calculate the quasiparticle (QP) band-structure for
crystalline GaAs in various states of electronic excitation by first order
perturbation theory \cite{GW2} from the real part of the self-energy: 
\begin{eqnarray}
E_{n{\bf k}}^{QP}=
E_{n{\bf k}}+\langle n{\bf k}|
Re\Sigma(E_{n{\bf k}}^{QP})-V_{xc}|n{\bf k}\rangle, 
\end{eqnarray}
with $Re\Sigma \equiv \frac{1}{2}(\Sigma^r+\Sigma^a)$. 
Using $\langle n{\bf k}|Re\Sigma|n{\bf k}\rangle=
Re\langle n{\bf k}|\Sigma^r|n{\bf k}\rangle$, (18) is solved by 
evaluating the real part of the retarded
self-energy, (15) and (16), for various distribution functions $f$.

The ``non-interacting'' mean field wavefunctions $|n{\bf k}\rangle$ and
eigenvalues $E_{n{\bf k}}$ as well as the
exchange-correlation potential $V_{xc}$ are obtained
within constrained LDA performed at the experimental lattice
constant. We use ${\it ab ~initio}$ pseudopotentials generated
with the scheme of Trouiller and Martins \cite{TM}. A plane wave basis
with an energy cutoff of 30 Ry is used to represent the Bloch
states. We use 50 valence plus conduction bands and discrete
Monkhorst-Pack meshes \cite{Monk} of ${\bf q}$ points (8x8x8) in the
first BZ in the sums of (15) and (16), while crystalline local-field 
effects in the dielectric matrix were included by summing over {\bf
G}, {\bf G'} up to a cutoff of 6 Ry. The integral over $\omega$ 
in (16) is evaluated by setting the integration limits to $\pm 50$ eV.

\begin{figure}
\epsfig{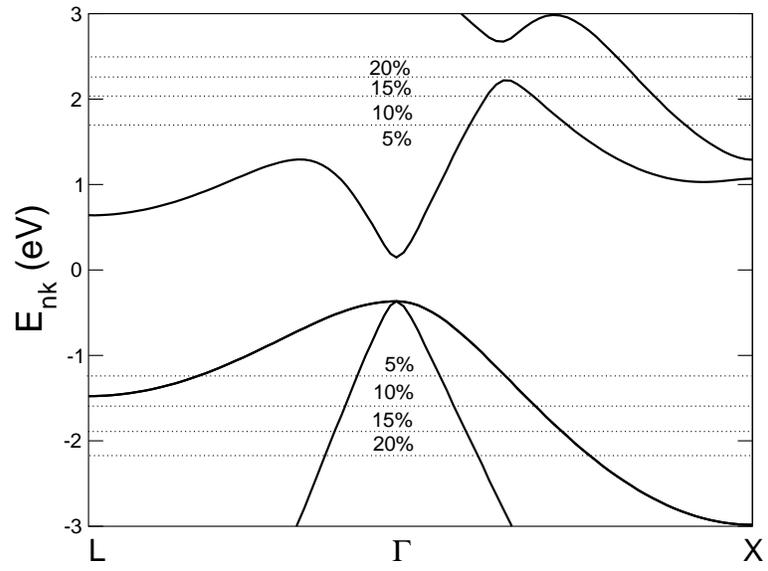}
\caption{Excited carrier occupation in GaAs; the LDA band structure shown here
is taken from the unexcited case.}
\label{Fig02}
\end{figure}

We consider five occupation number distributions $f$, each
corresponding to a different excited carrier density: 0\%, 5\%, 10\%,
15\% and 20\% of the total number of valence electrons in the system
(the coarseness of our meshes of ${\bf q}$ points precludes us from
considering excited carrier densities between 0\% and 5\%). Each is
characterized by a quasi-Fermi level of the conduction bands: the
energy below which all conduction states are filled. The energy above
which all valence states are empty is then determined by requiring charge
neutrality. Note that our assumptions imply that the individual
excited electron and hole populations \cite{occ} are in quasiequilibrium at 
T=0. Fig. \ref{Fig02} shows the excited carrier occupation in the 
Brillouin zone (BZ), for the various excited carrier densities
considered in this work. Because the effective mass of the conduction
band at $\Gamma$ is very small, the $\Gamma$ valley of the conduction
band can accommodate less than 1\% of the total number of valence
electrons ($n_{tot}$). It follows that when the excited carrier
density is several percent of $n_{tot}$, the excited electrons must
reside in a combination of conduction $\Gamma$, L and X valleys. It is 
these high carrier densities that we are concerned with here, given the 
conditions reached in recent pump-probe experiments \cite{Mazur1}. 

\section{Results and Discussion}

\begin{figure}
\epsfig{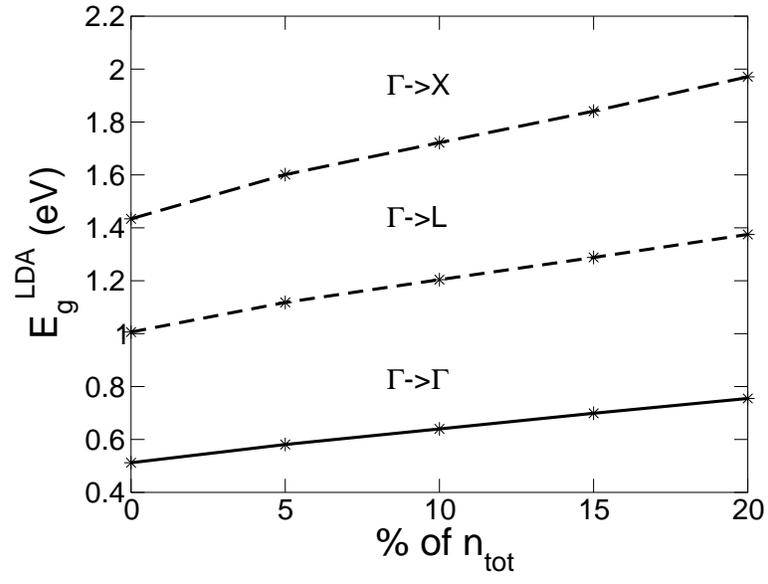}
\caption{Theoretical Kohn-Sham band gaps of GaAs calculated within the
constrained LDA.}
\label{Fig03}
\end{figure}

\begin{figure}
\epsfig{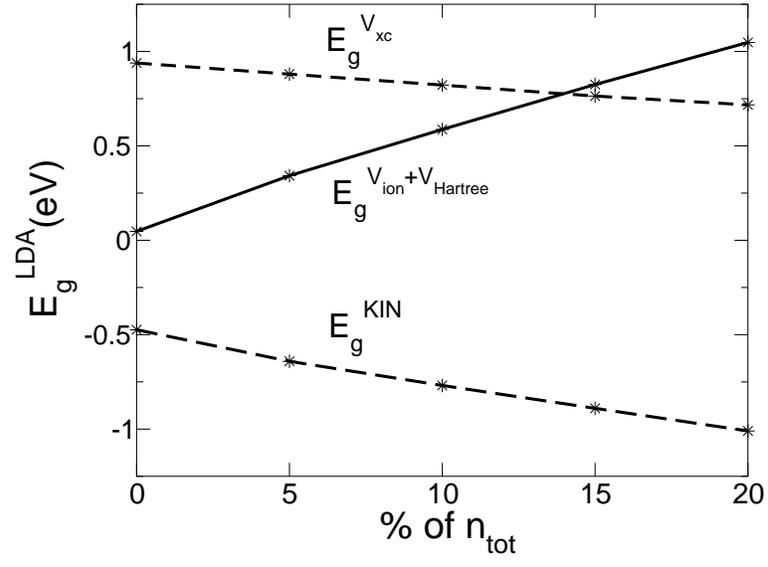}
\caption{Different contributions to Kohn-Sham band gap at $\Gamma$ for
GaAs calculated in the constrained LDA.}
\label{Fig04}
\end{figure}

Before presenting the results of our calculations of the QP self-energy 
operator, we discuss the changes in the band gap caused by the excitation 
at the mean-field (in this case, LDA) level. 
Fig. \ref{Fig03} presents the constrained LDA calculation results of
BGR. We see that the valence-conduction gap $E_g^{LDA}$ increases linearly with
the excited carrier density, the slope being only slightly smaller 
for the direct gap
at $\Gamma$ than for the indirect gaps at L and X. In order to
understand these results, we analyze the dependence
of the direct gap at $\Gamma$ on the excited carrier
density. We do this by decomposing the Kohn-Sham hamiltonian into the
kinetic energy part $K$, the ionic potential plus the
Hartree potential $V_{ion}+V_{Hartree}$, and the LDA
exchange-correlation potential $V_{xc}$. Their 
individual contributions to BGR are presented in 
Fig. \ref{Fig04}, which shows that the dominant term leading to the increase
of the constrained LDA valence-conduction gap at $\Gamma$ is the Hartree contribution.
Why is the Hartree term in the band gap increasing with the excited carrier
density? The answer is straightforward if we take into account that:
1) high-lying valence electrons are localized more near As, while 
low-lying conduction
electrons are localized more near Ga. 2) under excitation the electron charge density 
goes from As to Ga. Then it follows that under excitation, the electrostatic 
energy between one electron and the total charge density goes down for valence
electrons and up for conduction electrons, resulting in an increase in
the valence-conduction band gap. We note that at the mean-field level, our 
results are different from those obtained by Kim et al. \cite{Kim} using a 
self-consistent screened empirical pseudopotential treatment. They found 
(for $10\%$ excited carrier density) a significant decrease 
of the indirect gaps ($\sim 1~eV$ for the $\Gamma$-X gap) and no 
change of the direct gap at $\Gamma$. We ascribe this to their neglect of 
the change in the Hartree term. Though some of this mean-field effect is 
undoubtedly included through the screening of their pseudopotentials 
by the excited carriers, we suspect that a large portion of this Hartree 
interaction is incorrectly accounted for in their approach. 

\begin{figure}
\epsfig{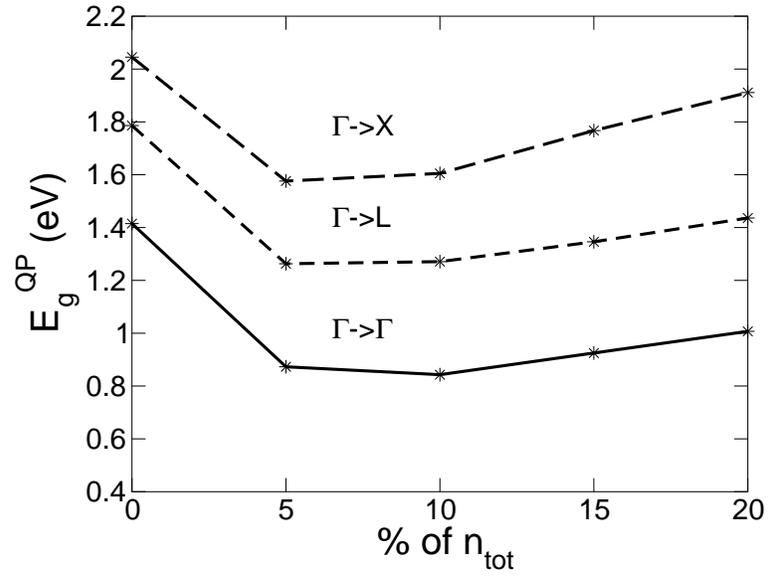}
\caption{The quasiparticle band gaps of GaAs as a function of excited
carrier density calculated within the $GW$ approximation.}
\label{Fig05}
\end{figure}

We now move to results at the QP level, by including the self-energy 
corrections in the $GW$ approximation. Fig. \ref{Fig05} presents the main 
results of our work, showing that the BGR picture changes dramatically
when we go from the mean-field level (constrained LDA) to the QP
level: we have a substantial decrease of all band gaps for relatively
low excited carrier density (less than 5\%), then the band gaps are rather
constant for excited carrier densities between 5\% and 10\%, while for
relatively high excited carrier densities (larger than 10\%) the band gaps
increase linearly. We note that the lowest value for the renormalized gap is about $0.8$ eV, which 
results from roughly a 10\% excited carrier density and is still a
direct gap at $\Gamma$.

\begin{figure}
\epsfig{file=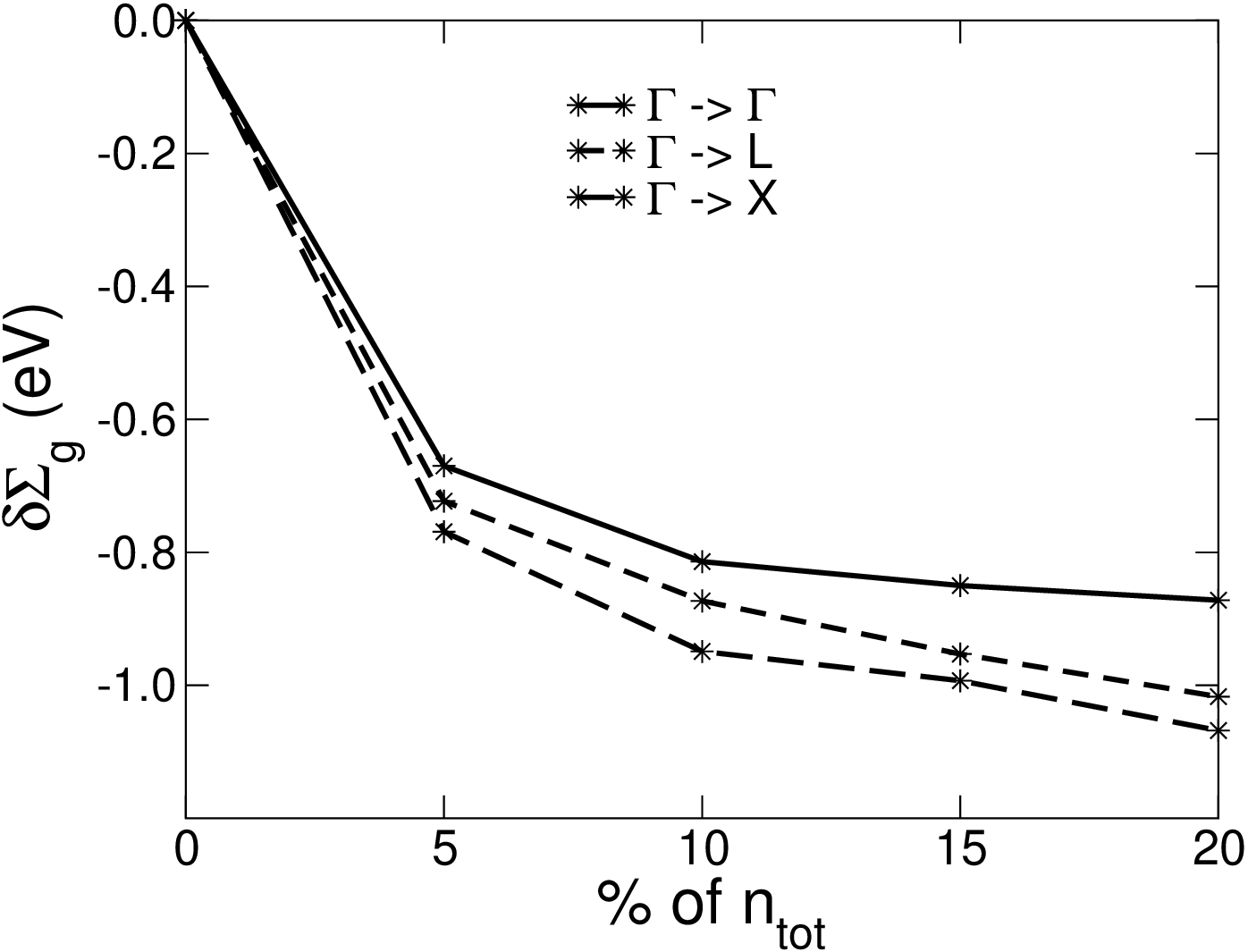, width=10.0cm}
\caption{Self-energy contribution to the band gap of GaAs as a
function of excited carrier density.}
\label{Fig06}
\end{figure}

We analyze the results of Fig. \ref{Fig05} by separating out
the contribution from the exchange-correlation self-energy part. 
We do that by introducing
$\delta\Sigma_g(n\%) \equiv \Sigma_g(n\%)-\Sigma_g(0\%) \nonumber$
(with 
$\Sigma_g \equiv
\langle c{\bf k}|Re\Sigma|c{\bf k}\rangle -
\langle v{\bf k}|Re\Sigma|v{\bf k}\rangle \nonumber $, $v$ and $c$ denoting
the valence and conduction bands near the Fermi level, {\bf k} being
the $\Gamma$, L or X)
and we interpret $\delta\Sigma_g$ as the self-energy contribution
to BGR. Fig. \ref{Fig06} shows the dependence of $\delta\Sigma_g$ with the
excited carrier density. We see that $\delta\Sigma_g$ is negative and roughly independent 
of the gap we are looking at (it is largest for the indirect
gap from $\Gamma$-X), and that it decreases rapidly with the
excited carrier density for relatively low densities. For higher excited 
carrier densities, it becomes rather independent of density. It is worth noting
that Kim et al. found a similar self-energy contribution for an 
excited carrier density of 10\%: 
$\delta\Sigma_g(10\%)\approx -0.9~eV$ (an average along the entire
$\Gamma$-X direction), with the largest value at X.

\begin{figure}
\epsfig{file=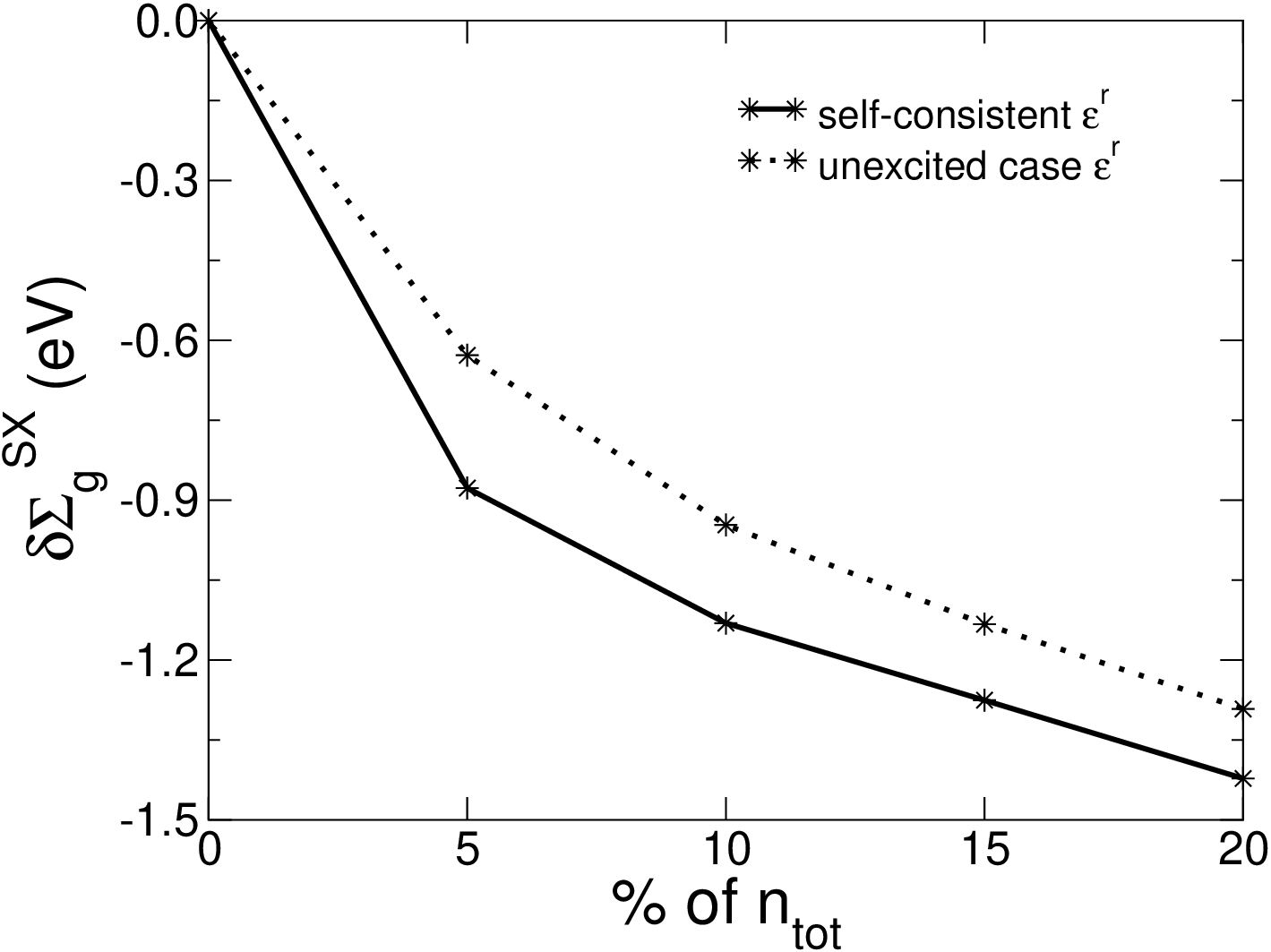, width=10.0cm}
\caption{Screened-exchange self-energy contribution to the band gap of
GaAs at $\Gamma$ as a function of excited carrier density.}
\label{Fig07}
\end{figure}

There are two main sources for the self-energy contribution
to BGR. One comes from the change in the Fermion occupation numbers, $f$, 
appearing  explicitly in eq. (15).
The other comes from the change in the electronic screening which 
manifests in both eqs. (15) and (16) through the dielectric 
function $\epsilon$ (of course, $\epsilon$ depends implicitly on $f$).
By increasing the excited carrier density, the screening acquires a metallic
character due to the increased number of 'free' electrons in the 
conduction band and 'free' holes in the valence band. We find that
the change in the electronic screening (affecting both the
screened-exchange and the Coulomb-hole terms) accounts only for a
small fraction of the BGR. For the excited carrier densities considered
in this work (in excess of $3.6 \times 10^{21}$ $\frac{e^{-}}{cm^{3}}$) the
major source for the self-energy contribution to
BGR was found to be the change in the Fermion
occupation numbers which enter the screened-exchange
self-energy expression (see the $f$ in eq.[15]). This picture may
not hold for excited carrier densities lower than $10^{21}$ $\frac{e^{-}}{cm^{3}}$
\cite{Godby}. We illustrate this observation in Fig. \ref{Fig07} 
where we show the screened-exchange self-energy contribution to
the direct gap at $\Gamma$ calculated in two ways: the solid line 
represents self-consistent calculations based on eq. (15), while 
the dotted line represents calculations based also on eq. (15), but
in which the dielectric function $\epsilon^r$ was computed for 
the unexcited case. Finally, we note that our results for
$\delta\Sigma^{SX}_g$ follow an approximate ${\frac{1}{3}}$ power law 
behavior in the excited carrier density; the same power law behavior for
$\delta\Sigma^{SX}_g$ can be obtained employing a very simple model based
on the parabolic band approximation and replacing $\epsilon^r$ with a
simple dielectric constant \cite{SR,Kim}.

Summarizing our results for BGR, we conclude that for relatively low excited
carrier densities (up to 10\% of the total valence electron density)
the QP valence-conduction band gap decreases (but never closes) primarily
due to the change in the Fermion occupation numbers entering
the screened-exchange QP self-energy expression, while for
higher excited densities the gap reaches a minimum value and undergoes
an upturn with increasing density. The increase is mainly due to the change in the 
electrostatic energy at the mean-field level described by the Hartree term. 
Our prediction that there is no band gap closure due to these 
valence band top $\longrightarrow$ conduction band bottom excitations 
agrees with the measurements of L. Huang et al. \cite{Mazur1}, in 
which the dielectric function of GaAs was measured at short ($\sim$ 100 fs) 
time delays of a pump-probe experiment and no signature of 
metallicity was found \cite{excuse_NT}. We also mention that the approximate 
${\bf k}$-independence of the BGR that we predict suggests that the overall 
shape of the dielectric function (i.e.- the distribution of oscillator 
strength in the optical absorption spectrum) in these excited configurations 
will be similar to that predicted by one of us \cite{Benedict} 
using an approach in which BGR 
was neglected. Finally, we add that our computations of the imaginary part of 
the QP self-energy for the cases considered above yield values small enough 
for us to conclude that lifetime broadening due to the scattering of excited 
carriers should be of little importance in determining the shape of the 
optical spectra.  

\section{Conclusions}
We have studied band gap renormalization in laser-excited 
GaAs. The cases we considered were those of intense excitation 
($\sim$ 10$\%$) in which excited electrons 
occupy the lowest-lying conduction states and the excited holes occupy the 
highest-lying valence states. 
The quasiparticle self-energy was computed with a non-equilibrium variant 
of the $GW$ approximation using the Keldysh technique.
Our study indicates that it is not possible to induce
complete  band gap closure by purely electronic
means. In reaching this conclusion, we found that the contribution of 
the Hartree term to BGR, describing the mean-field interaction between an excited 
electron (or hole) and the charge density of the remaining electrons, is 
important and cannot be neglected. Our findings seem to support experimental 
observations \cite{Mazur1} in which no evidence of 
electronically-induced band gap closure was seen. 

\section{Acknowledgments}
C.D.S. would like to thank Dr. David J. Roundy for helpful discussions
regarding the constrained LDA calculations. We also thank Dr. Eric K. 
Chang for his contributions to the early stages of this project. 
This work was supported by the NSF under Grant No. DMR0087088, and the
Office of Energy Research, Office of  Basic Energy Sciences, Materials 
Sciences Division of the U.S. Department  of Energy (DOE) under  Contract No. 
DE-AC03-76SF00098. Computer time was provided by the DOE at the 
Lawrence Berkeley National Laboratory (LBNL)'s NERSC center. Portions
of this work were performed under the auspices of the DOE
by University of California Lawrence Livermore National Laboratory
(LLNL) under contract No. W-7405-Eng-48. Collaborations 
between LLNL and LBNL were facilitated by the DOE's
Computational Materials Sciences Network.

\renewcommand{\theequation}{\theappendix\arabic{equation}}
\setcounter{equation}{0}
\section*{Appendix}
In this section, we present the Langreth rule \cite{Langreth} for 
analytic continuation of {\it products} of functions integrated over contour 
C. This is necessary when evaluating the $GW$ self-energy expression.   
The functions introduced in (6) satisfy the Keldysh relation:
\begin{eqnarray} 
A^r(t,t') - A^a(t,t') = A^>(t,t')- A^<(t,t'),
\end{eqnarray}
and also that for any functions $A$ and $B$:
\begin{eqnarray} 
A^r(t,t') B^a(t,t') = A^a(t,t') B^r(t,t') = 0.
\end{eqnarray}

The Langreth rule for analytic continuation states that if on the contour:
\begin{eqnarray}
F=\int_C AB \nonumber
\end{eqnarray}
then on the real time axis (for a concise derivation see \cite{Haug}):
\begin{eqnarray}
F^r=\int_{-\infty}^{\infty} A^rB^r \nonumber \\
F^a=\int_{-\infty}^{\infty} A^aB^a \nonumber \\
F^<=\int_{-\infty}^{\infty} A^rB^< + A^<B^a \nonumber \\
F^>=\int_{-\infty}^{\infty} A^rB^> + A^>B^{a}. 
\end{eqnarray}

It can also be shown that if:
\begin{eqnarray}
F(\tau,\tau')=A(\tau,\tau')B(\tau,\tau') \nonumber \\
D(\tau,\tau')=A(\tau,\tau')B(\tau',\tau) \nonumber,
\end{eqnarray}
then:
\begin{eqnarray}
F^r(t,t')=A^<(t,t')B^r(t,t')+A^r(t,t')B^<(t,t')+A^r(t,t')B^r(t,t'),
\end{eqnarray}
and:
\begin{eqnarray}
D^r(t,t')=A^<(t,t')B^a(t',t)+A^r(t,t')B^<(t',t)\nonumber \\
D^a(t,t')=A^<(t,t')B^r(t',t)+A^a(t,t')B^<(t',t)\nonumber \\
D^<(t,t')=A^<(t,t')B^>(t',t) \nonumber \\
D^>(t,t')=A^>(t,t')B^<(t',t).
\end{eqnarray}
This result can be directly applied to the evaluation of the $GW$ self-energy 
expression, where $G$ and $W$ replace $A$ and $B$ respectively (as well 
as to the polarizability $P$, replacing $A$ and $B$ by $G$).

\clearpage

\clearpage

\clearpage

\clearpage

\clearpage

\clearpage

\clearpage

\begin{references}
\bibitem{excuse_single}
The physics of electron-hole interaction is left out of this treatment 
since we focus our attention on the one-particle Green function. For 
a discussion of such two-particle effects in highly excited GaAs, see 
\cite{Benedict}.
\bibitem{excuse_NT}
At larger times, changes in the dielectric 
function were observed, and are believed to be signatures of a non-thermal 
structural transformation to a metallic amorphous state.
\end{references}
\end{document}